\documentclass[a4paper,aps,prl,twocolumn,superscriptaddress,amsmath,amssymb]{revtex4}
\usepackage{hyperref}
\usepackage{graphicx,color}

\newcommand{\dd}{\mathrm{d}}

\newcommand{\mean}[1]{\langle #1 \rangle}
\newcommand{\Int}[1]{\int\dd #1\;}

\renewcommand{\vec}[1]{\mathbf #1}

\newcommand{\mrm}[1]{\mathrm {#1}}

\newcommand{\eps}{\varepsilon}

\newcommand{\sig}{\sigma}

\newcommand{\x}{\vec r}

\newcommand{\nois}{\boldsymbol\zeta}

\newcommand{\ps}{\Psi_\text{s}}
\newcommand{\vel}{\overline{v}}

\newcommand{\mvisc}{\eta_\mathrm{eff}}

\begin{document}

\title{Discontinuous Thinning in Active Microrheology of Soft Complex Matter}

\author{R. Wulfert}
\author{U. Seifert}
\affiliation{II. Institut f\"ur Theoretische Physik, Universit\"at Stuttgart,
  Pfaffenwaldring 57, 70550 Stuttgart, Germany}
\author{T. Speck}
\affiliation{Institut f\"ur Physik, Johannes Gutenberg-Universit\"at Mainz,
  Staudingerweg 7-9, 55128 Mainz, Germany}

\begin{abstract}
  Employing theory and numerical simulations, we demonstrate discontinuous force thinning due to the motion of an external probe in a host medium, which we approximate as structureless. When the driving of the probe exceeds a critical force, the microviscosity of the medium drops abruptly by about an order of magnitude. This phenomenon occurs for strong attractive interactions between probe and a sufficiently dense host medium.
\end{abstract}


\maketitle


Active microrheology (AMR) is an experimental technique to determine local rheological properties of a ``soft'' material by measuring actively manipulated probes (for reviews see, \emph{e.g.}, Refs.~\citenum{squi10,wils11,puer14}). For colloidal suspensions, it has been employed close to the glass transition by dragging a magnetic bead~\cite{habd04}, for hard spheres over a wide range of volume fractions~\cite{wils09}, and to force the local melting of a colloidal crystal~\cite{dull11,weeb12}. It has also become an important technique for the study of mechanical properties of biological matter~\cite{mizu08,wilh08} and fluid interfaces~\cite{choi11}. Moreover, it allows quantitative insights into the mechanical response of glassy materials~\cite{jack08,gazu09,wint12,schr13} and crowded systems~\cite{beni13}.

An established theoretical model for AMR is the ``simple paradigm'' of Squires and Brady~\cite{squi05,carp05}, which has been applied predominantly to hard-sphere suspensions. It allows to describe a broad class of materials governed by short-ranged repulsions through a mapping onto hard spheres via an effective diameter. Conventional macrorheology deals with averages of quantities like stress and strain, which requires sample materials to be both sufficiently homogeneous and spatially extended. In AMR, however, the host medium is strained only in the vicinity of the moving probe. It is this locality of the induced flow fields that makes AMR so useful for the exploration of media that are either confined (\emph{e.g.} in living cells), heterogeneous on mesoscopic length scales, or just difficult to procure in larger quantities. Like macrorheology, AMR centers around measuring and understanding the non-Newtonian behavior of complex fluids, albeit from a more introversive angle. This predominantly concerns their relaxation in response to perturbations~\cite{zia13,gome15} as well as their thinning and thickening behavior under steady flows~\cite{gome14,wang15}. To this end, the simple paradigm relates changes in the microstructure to microrheological properties.

Arguably the most extreme form of non-linear behavior manifests itself in materials with discontinuous flow curves. Such discontinuities give rise to rather spectacular effects, a paradigmatic example being the sudden solidification of cornstarch suspensions above some critical strain rate. Although the phenomenon of discontinuous shear thickening~\cite{brow14a} has long been known, its underlying mechanism remains the subject of ongoing theoretical and experimental investigations~\cite{seto13,wyar14,fall15}. The opposite case, discontinuous shear thinning, has received far less attention and so far has only been reported in connection with order-disorder transitions under shear flow~\cite{chen90,rigo92}.

Here we report the discontinuous thinning of a soft material in response to the motion of a forced probe \emph{attracting} the surrounding host material. A numerical analysis based on an extension of the simple paradigm formalism~\cite{squi05} reveals a dynamically unstable regime within the velocity-force relation of the driven probe. To this end we discuss how a slight modification of the simple paradigm justifies its application to a (colloidal) bath of arbitrary number density as long as the bath particles are small enough not to be structurally correlated too strongly. We corroborate our results with Brownian dynamics simulations showing that, for sufficient density and attraction strength, crossing a certain threshold of the applied driving force induces a marked instantaneous drop in the microviscosity of the host medium. Moreover, for finite-time force protocols across this discontinuity, the dynamic lag of the microstructure causes hysteretic behavior, showing up as a loop in the velocity-force relation. The underlying mechanisms of this discontinuous force-thinning turn out to be rather generic and intuitive. It can be accounted for solely by considering the perturbed microstructure of bath particles relative to the probe drifting at different speeds.


We model the host medium as $N$ bath particles suspended in a solvent and moving in a periodic box of volume $V$. With a hydrodynamic diameter $b$ and number density $\rho\equiv N/V$, the particles account for a volume fraction of $\phi\equiv(\pi/6)b^3\rho$. A spherical probe of diameter $a$ is pulled through the bath by a constant external force $f\vec{e}_x$. Interparticle forces $\vec F_k=-\nabla_kU$ are derived from the superposition $U=U(\{\vec{x}_k\})$ of pair potentials $u(r)$ and $u_b(r)$, acting between the probe and bath particles and among bath particles, respectively. Throughout, we measure lengths in units of $(a{+}b)$, energies in units of $k_\mrm{B}T$, and diffusivity in $(D_a{+}D_b)$, which is the relative diffusivity between the probe and a single bath particle. The solvent is assumed to be Newtonian with a viscosity $\eta$ large enough to render the colloidal inertia irrelevant. The resulting Brownian motion generates particle trajectories $\vec{x}_k(t)$ described by the overdamped Langevin equations
\begin{equation}
  \label{eq:bd}
  \dot{\vec{x}}_k = \mu_k \left[\vec{F}_k + f\vec{e}_x \delta_{0k} \right] + \nois_k
\end{equation}
with $k=0$ referring to the probe particle. The noise $\nois_k(t)$ has zero mean and correlations $\mean{\nois_k(t) {\otimes} \nois_l(t')} = 2 \mu_k \delta_{kl} \mathbf{I} \delta(t-t')$ with identity matrix $\mathbf I$. Stokes' law of friction for spherical particles yields (in our units) the probe mobility $\mu_0=1+\alpha^{-1}$ and mobility $\mu_b=1+\alpha$ of bath particles with size ratio $\alpha\equiv a/b$.

An equivalent description is given in terms of the joint probability distribution $\Psi(\{\vec x_k\},t)$ with local mean velocities $\vec v_k=\mu_k(\vec F_k+f\vec e_x\delta_{0k}-\nabla_k\ln\Psi)$. After switching to bath coordinates relative to the probe, $\{\vec{r}_{k=1..N} \equiv \vec{x}_k - \vec{x}_0\}$, all gradients with respect to the absolute position $\vec{x}_0$ of the probe vanish due homogeneity of the system. Conservation of probability amounts to the many-body Smoluchowski equation, which in the stationary case reads
\begin{equation}
  \label{eq:mpSeq_NESS}
  \sum_{k=1}^N\nabla_k \cdot (\vec v_k-\vec v_0) \ps = 0.
\end{equation}
It determines the nonequilibrium steady state distribution $\Psi_\mrm{s}(\{\x_k\};f)$, which includes the microstructural deformations induced by the external driving. Unless otherwise stated, averages $\mean{\cdot}$ are taken with respect to $\Psi_\mrm{s}$.

\begin{figure*}[t!]
  \centering
  \includegraphics{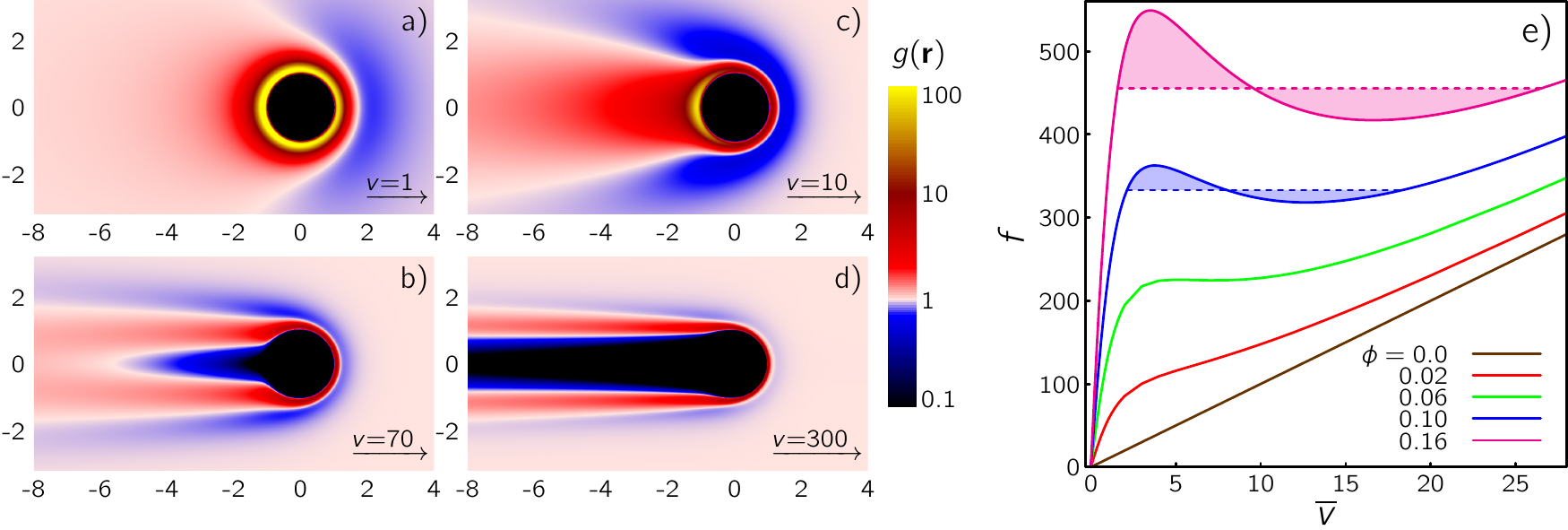}
  \caption{(a-d)~Microstructural deformation around the moving probe in terms of the conditional probability $g(\x;\vel)$ in the $(x,y)$-plane. For small mean velocities $\vel$, the probe is dragging along an excess of trailing particles, which in turn generate a considerable frictional load. With increasing flow around the probe, this excess is flushed out of the potential well and a depleted wake forms behind the probe, leading to a marked drop in the effective friction $\zeta$. (e)~Velocity-force relations $f(\vel)$ of an attractive probe for various volume-fractions $\phi$. Above $\phi\simeq0.06$, the flow curves become non-monotonous with regions where less driving force $f$ is required for the probe to move with a higher mean-velocity $\vel$, rendering the system dynamically unstable. Shaded areas hint at a tentative equal-area Maxwell construction to replace the unstable region.}
  \label{fig:instab}
\end{figure*}

For pairwise interactions the mean frictional drag
\begin{equation}
  \label{eq:zeta}
  \zeta(f) \equiv \vec e_x\cdot\mean{\vec F_0} = \rho\Int{\x} (\vec e_x\cdot\vec e_r) \, u'(|\x|) \, g(\x;f)
\end{equation}
exerted on the probe reduces to an integral involving the one-body density $\rho_1(\x;f)=\rho g(\x;f)$ of bath particles around the probe with conditional probability $g(\x;f)$ to find a bath particle at $\x$ given the probe at the origin. In general, this density will depend both on the driving strength $f$ and bath density $\rho$. The ensuing velocity-force relation
\begin{equation}
  \label{eq:vf}
  \vel(f) \equiv \vec e_x\cdot\mean{\vec v_0} = \mu_0[f-\zeta(f)]
\end{equation}
expresses the nonlinear response of the probe to the external driving, reducing the solution of the microrheological problem to the calculation of $g(\x;f)$.

The one-body density $\rho_1(\x)$ is the first member of an open hierarchy. In order to achieve closure, we approximate the two-body density $\rho_2(\x,\x')\approx \rho_1(\x)\rho_1(\x')$, which neglects \emph{any} correlations between bath particles. For strongly interacting particles, such an approximation would correspond to the dilute limit $\phi\to0$. However, here we use it as a model for structureless soft media at arbitrary density such as polymer melts with probe sizes larger than the microscopic correlation length~\cite{loui00}. Within this approximation, we obtain the pair Smoluchowski equation
\begin{equation}
  \label{eq:pSeq}
  \nabla \cdot \left[2\nabla + \vel\vec{e}_x + 2u'(|\x|) \vec e_r \right] g(\x;\vel) = 0,
\end{equation}
which is essentially the constitutive equation of the simple paradigm~\cite{squi05} including continuous pair interactions $u(r)$ between probe and bath particles. Eq.~\eqref{eq:pSeq} has the form of an advection-diffusion equation with $\vel$ quantifying the advection of bath particles as seen from the moving probe, which is of course equal in magnitude to the mean probe velocity in the laboratory frame. In the units chosen above, $\vel$ is the Peclet-number commonly used to quantify the relative strength of advection \emph{vs.} diffusivity~\cite{squi05}. Note that solving Eq.~(\ref{eq:pSeq}) yields $g(\x;\vel)$ as a function of the mean velocity $\vel$ \emph{independent} of the density $\rho$. Hence, we have $\zeta=\rho\hat\zeta(\vel)$ with force $f(\vel)=\rho\hat\zeta(\vel)+\vel(f)/\mu_0$ from Eq.~(\ref{eq:vf}) necessary to pull the probe with mean velocity $\vel$ against this drag. Whereas in the hard-sphere case an analytical solution to Eq.~\eqref{eq:pSeq} is available in the form of a series expansion~\cite{squi05}, here we have to resort to a numerical scheme. To solve Eq.~(\ref{eq:pSeq}), with the axial symmetry around the force direction taken into account, we employ spherical coordinates and expand $g(r,\theta)$ into Legendre polynomials. Discretizing the modes with respect to $r$ leads to a system of linear equations, which is solved by standard techniques. After having sampled the microrheological flow curve with a sufficient number of value pairs $(\vel,f)$, the $f(\vel)$ relation can be numerically interpolated and inverted to give the desired $\vel(f)$ velocity-force relation for given density $\rho$.


As a convenient pair potential combining attractions with a steep repulsive inner part mimicking hard-core exclusion, we choose the standard Lennard-Jones-(12,6) potential 
\begin{equation}
  \label{eq:lj}
  u(r) = 4 \epsilon \left[ \left(\frac{\sigma}{r}\right)^{12} - \left( \frac{\sigma}{r} \right)^6 \right]
\end{equation}
acting between probe and bath particles. In the case of a passive probe, \emph{i.e.} $f=0$, the equilibrium distribution $g_\text{eq}(r)=\exp[-u(r)]$ of bath particles is isotropic and Boltzmannian. Driving the probe with finite $f$ breaks isotropy and leads to a characteristic, primarily dipolar, deformation of the bath microstructure around the probe. For a rather high interaction strength $\eps=5$ and length scale $\sig=1$ of the potential, $g(\x)$ with increasing probe drift velocity $\vel$ is plotted in Fig.~\ref{fig:instab}(a-d). In the case of a slowly moving probe [Fig.~\ref{fig:instab}(a)] the potential well is highly populated with bath particles, while the region of maximal excess is shifted downstream due to advection. Naturally, having to drag along these trailing particles generates considerable frictional resistance against the probe drift. Upstream, suction towards the potential minimum creates a halo-like region of depleted bath density. With increasing drift velocity $\vel$, the directional bias in the microstructure becomes more and more pronounced, see Fig.~\ref{fig:instab}(b,c). As bath particles now have less time to react to the potential forces while passing the probe, its upstream range of influence contracts radially. Meanwhile, the probe trails behind it a wake of ever increasing spatial dilation. Although this wake is still carrying an excess of bath particles, its peak density is reduced. When the drift becomes strong enough, however, advection dominates over the attractive potential forces and the downstream excess starts being swept away, resulting in a more or less evacuated tail-like zone directly behind the probe, see Fig.~\ref{fig:instab}(c). Having cast off its ``baggage'', the probe can now propagate much more freely. It is only for very strong driving that the distribution of bath particles starts to resemble the typical hard-sphere picture [Fig.~\ref{fig:instab}(d)], where regardless of driving strength one observes a buildup of particles in front of the probe and a depleted wake trailing it.

The velocity-force relations corresponding to this intriguing, highly nonlinear behavior are depicted  in Fig.~\ref{fig:instab}(e) for $\alpha=19$ and for various volume-fractions $\phi$. Their representation as $f(\vel)$, as opposed to the inverse $\vel(f)$, is intended to emphasize that our numerical scheme is also reversed, in the sense that we calculate, one at a time, the forces $f$ necessary to drive the probe with given mean velocities $\vel$.
The most salient feature of the series of $f(\vel)$ curves in Fig.~\ref{fig:instab}(e) is that above a particular volume-fraction, here specifically $\phi \simeq 0.06$, they become non-monotonous over an intermediate $\vel$-range. It implies the existence of conditions for which the probe requires less driving in order to travel faster, constituting a dynamically unstable regime. At the same time, the function $f(\vel)$ is no longer globally invertible. Since the force $f$ is the physical control parameter, and stochastic dynamics guarantees a unique steady-state with an unambiguous drift velocity $\vel$, this implies a discontinuous change of the velocity-force relation $\vel(f)$. Such a behavior of an order parameter is well known in conventional thermodynamics, where it signals a phase transition. For example, in the van der Waals theory the volume as a function of pressure shows a similar loop due to the competition between energy (favoring the dense phase) and entropy (favoring the dilute phase). The Maxwell construction removes the loop by equating the free energies of both phases. We note the analogy between intensive pressure and force, and extensive volume and the distance traveled by the probe. As a for now purely tentative measure to restore invertibility, we propose, in the spirit of a Maxwell construction, that the unstable sinusoidal part of $f(\vel)$ ought to be replaced by a horizontal line $f=\text{const.}$ according to an equal-area rule shown as shaded areas in Fig.~\ref{fig:instab}(e).


In order to assess the physical validity of the resulting flow curves, we have performed Brownian dynamics (BD) simulations. In contrast to solving Eq.~(\ref{eq:pSeq}), we now control the driving force $f$, allowing us to sample the true $\vel(f)$ relations over the critical region. There is, however, one important caveat: In simulations it is not possible to completely neglect bath correlations since there will be indirect correlations mediated by the probe even for non-interacting bath particles with $u_b(r)=0$. However, it turns out that in the case of large size ratios $\alpha$ these correlations becomes negligible and the two approaches converge. In the following, we choose $\alpha=19$, \emph{i.e.}, the hydrodynamic radius of the probe is 19 times larger than that of the ideal bath particles.

\begin{figure}[t!]
  \centering
  \includegraphics{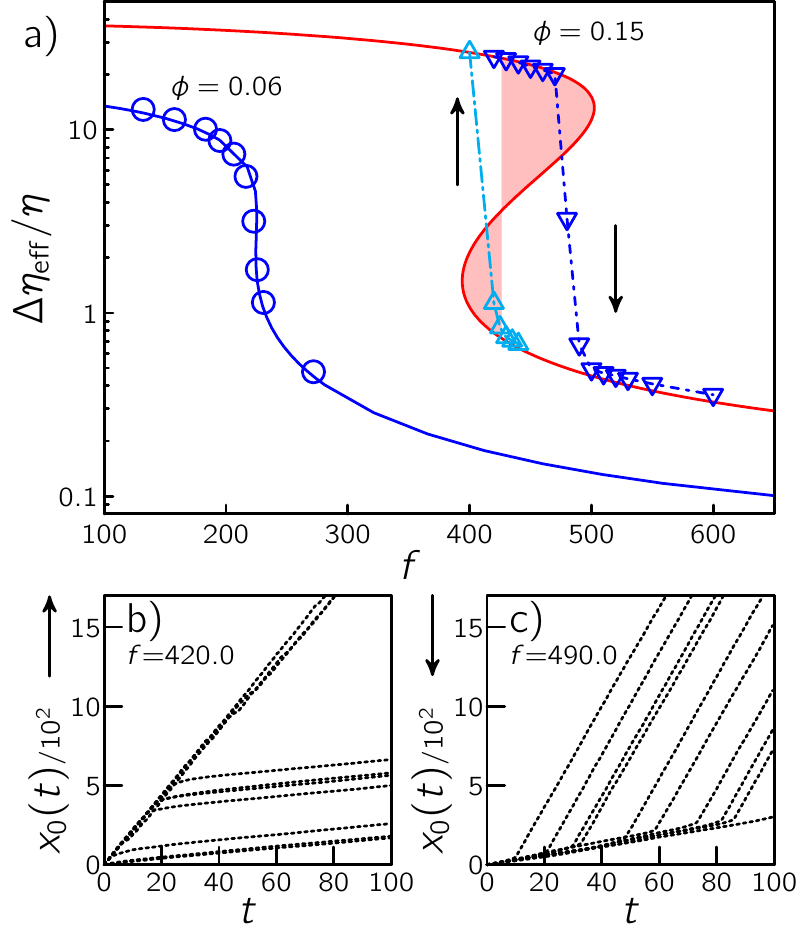}
  \caption{(a)~Relative microviscosity increment $\Delta\eta_\text{eff}/\eta = \mu_0f/\vel-1$ as a function of driving force $f$ for various bath volume fractions $\phi$. Solid lines represent numerical solutions of the pair Smoluchowski-Eq.~\eqref{eq:pSeq}. The data points obtained from BD simulations match the numerical solutions quite well, with a statistical error smaller than the symbol size. For $\phi\simeq 0.15$ the simulation data reveal a sharp discontinuous transition between two separate frictional regimes. The force at which the transition occurs depends on the rate and direction along which one varies $f$, as indicated by the arrows, bringing about hysteretic behaviour. (b,c)~Probe positions $x_0(t)$ as a function of time obtained from single simulation trajectories for (b)~initial forces below and (c)~above the discontinuity. In order to escape the metastable initial state, a rare fluctuation has to carry the system close enough towards its stable steady state, followed by a marked and almost instantaneous jump of the probe velocity.}
 \label{fig:disc_thinning}
\end{figure}

For $N=32000$ particles plus the probe, trajectories $\vec{x}_k(t)$ are integrated according to a time-discretized Euler-scheme based on Eq.~\eqref{eq:bd} with a timestep $\Delta t = 5\times 10^{-5}$. A preliminary relaxation time $\tau_\text{rel}=50$ is reserved for the system to reach its nonequilibrium steady-state. An estimate of the mean probe velocity $\vel(f)$ is then obtained as an average over $10$ statistically independent runs per given force $f$, corresponding to an average $\mean{\cdot}$ with respect to the noise $\{\nois_k(t)\}$. Each run has a duration of $\tau_\text{sim}=200$, in which the relative distance of a passive probe and a single bath particle would increase diffusively by roughly $400$-times their hydrodynamic contact distance $(a{+}b)/2$.

Introducing microrheological terminology, we consider the relative microviscosity $\mvisc/\eta=\mu_0f/\vel$. Its definition hypothesizes a Newtonian medium of the same viscosity, whose purely Stokesian resistance to the probe equals the total friction $(\vel/\mu_0+\zeta)$ of our suspension.
In contrast to a Newtonian viscosity, however, the microviscosity $\mvisc$ of complex fluids will generally be a function of $\vel$ or $f$, respectively, just as their macroscopic shear-viscosity will be a function of strain-rate or stress. In order to elicit the nonlinear contribution of the colloidal bath to the microviscosity, Fig.~\ref{fig:disc_thinning} shows the relative increment $\Delta\mvisc/\eta=\mvisc/\eta-1$ of the microviscosity as a function of driving force $f$ for both the BD simulations and the solution of Eq.~\eqref{eq:pSeq}. For the lower volume-fraction $\phi=0.06$, where the velocity-force relation is still unambiguous, the two methods match almost perfectly, justifying in retrospect the $u_b(r)=0$ approach in adapting the simulated system to the pair Smoluchowski model. They both reveal pronounced but continuous thinning over an intermediate force-range, with the relative microviscosity increment declining more than one order of magnitude, followed by an asymptotic decay towards zero. This result is in stark contrast to the behavior of a hard-sphere suspension, where in the limit $f\to\infty$ the microviscosity asymptotically converges to a quasi-Newtonian plateau with a value of $\Delta\mvisc(f\to\infty)=\Delta\mvisc(f\to0)/2$~\cite{squi05}. This means that for strong driving the bath contribution $\zeta$ to the total friction $(\vel/\mu_0+\zeta)$ becomes more and more irrelevant.

We now study the thinning behavior for volume-fractions above $\phi=0.06$, where the instability develops. For $\phi=0.15$, BD simulations reveal a sharp discontinuous transition between two frictional regimes within the unstable region. The microviscosity increment $\Delta\eta_\text{eff}(f)$ suddenly drops (or surges) by more than one order of magnitude, depending on the direction along which $f$ is varied. The fact that the simulation results in Fig.~\ref{fig:disc_thinning} fork into two branches for $\phi=0.15$ is due to the rather long crossover time from one metastable state to the other. Hence there is a difference whether the initial relaxation of the system starts with $f=0$ and the force being slowly ramped up, see \ref{fig:disc_thinning}(c); or whether it starts with a force beyond the unstable regime, which is then gradually reduced towards its steady-state value, see Fig.~\ref{fig:disc_thinning}(b). For finite relaxation times, this leads to hysteretic behavior around a critical force estimated from the Maxwell-construction, which again is shown as a shaded area in Fig.~\ref{fig:disc_thinning}.


In conclusion, we have combined an extension of the simple paradigm with simulations in order to demonstrate both the existence and origin of discontinuous thinning of a soft environment embedding a forced, attractive probe. The underlying mechanism rests on the aggregation of the host medium in the potential well and thus depends both on density and attraction strength. Specific details of the pair potential, however, are not essential to the phenomenon itself. In that respect it appears to be quite robust and generic. Considering a structureless medium has allowed us to clearly expose the principal physical contributions, but corrections could be incorporated using, \emph{e.g.}, a DFT-like approach~\cite{rein13,rein14} in the closure of the Smoluchowski equation~(\ref{eq:mpSeq_NESS}). On the practical side, our results show that it is paramount to carefully take into account the interactions between probe and medium in interpreting experimental data obtained through active microrheology in terms of simple theoretical models.


\acknowledgments

We acknowledge financial support by the DFG (grant numbers SE1119/3-2 and SP1382/1-2).


\end{document}